\documentclass[11pt]{article}

\usepackage{doublespace,cite}
\makeatletter
\renewcommand{\@cite}[2]{\textsuperscript{(#1)}}
\makeatother
 
\author{
Barbara Niethammer\textsuperscript{1}
\quad and \ Robert L. Pego\textsuperscript{2} 
}

\title{ 
Non-self-similar behavior in the \\
LSW theory of Ostwald ripening}

\date{December 1998}

\newcommand{\nwc}{\newcommand}

\newtheorem{prop}{Proposition}[section]
\newtheorem{lemma}[prop]{Lemma}

\newtheorem{theorem}[prop]{Theorem}
\newtheorem{corollary}[prop]{Corollary}
\newtheorem{definition}[prop]{Definition}
\newenvironment
  {proof}{\noindent {\bf Proof:}}{\bigskip}
 
\newcommand {\eps}{\varepsilon}
\nwc{\D}{\partial}
\nwc{\vdot}{\Lambda}
\nwc{\F}{\varphi}
\nwc{\osc}{\mathop{\rm osc}}

\nwc{\Fname}{flipped}
\nwc{\y}{y}
\nwc{\z}{z}
\nwc{\ka}{\kappa}
\nwc{\rh}{\varrho}
\nwc{\chv}{{\cal V}}
\nwc{\chu}{{\cal U}}
\nwc{\chy}{{\cal Y}}
\nwc{\ty}{{\tilde y}}
\nwc{\tchy}{\tilde{\chy}}
\nwc{\ttau}{{\tilde\tau}}
\nwc{\preSF}{{\tilde\psi}}
\nwc{\SF}{\psi_*}
\nwc{\SFP}{\Psi_p}
\nwc{\SFI}{\Psi_\infty}
\nwc{\ks}{\kappa_*}
\nwc{\rs}{\rho_*}
\nwc{\lnp}{\rho}
\nwc{\psii}{\psi_\infty}
\nwc{\kai}{\kappa_\infty}
\nwc{\mf}{\varpi}
\nwc{\GP}{G_p}
\nwc{\KK}{{\bar K}}
\nwc{\CB}{{\bar C}}
\nwc{\Fi}{{F_\infty}}
\nwc{\supp}{\mathop{\rm supp}\nolimits}

\usepackage{amsfonts}
\nwc{\R}{{\mathbb R}}
\nwc{\third}{{1/3}}
\nwc{\RCD}{rcd([0,1])}
\nwc{\rcd}{rcd}
\nwc{\BDD}{bdd([0,1])}
\nwc{\Psp}{{\cal P}_0}  
\nwc{\loc}{{\rm \scriptstyle loc}}

\nwc{\ds}{\displaystyle}
\nwc{\eqref}[1]{(\ref{#1})}
\newcommand {\bedis} {\begin{displaymath}}
\newcommand {\edis} {\end{displaymath}}
\newcommand{\newbeqna} {\renewcommand {\arraystretch} {2}
                        \begin {displaymath} \begin {array}{crcl}}
\newcommand{\neweqna}{\end{array} \end {displaymath}}
 
\newcommand{\fbeqna}{\renewcommand {\arraystretch} {1.3}
\begin {displaymath}\begin{array}{rcll}}
\newcommand{\feqna}{\end{array}\end{displaymath}}
 
\newcommand {\beqna} {\begin{eqnarray*}}
\newcommand {\eqna} {\end{eqnarray*}}
\newcommand {\beqn} {\begin{eqnarray}}
\newcommand {\eqn} {\end{eqnarray}}
\newcommand {\be} {\begin{equation}}
\newcommand {\ee}{\end{equation}}

\nwc{\sfrac}[2]{\frac{\scriptstyle #1}{\scriptstyle #2}}

\def\drawbox#1#2{
   \hbox{\vrule\vbox{\hrule width#1\vskip #2\hrule width#1}\vrule}}
\nwc{\eproof}{\drawbox{.1in}{.1in}} 

\begin{document}

\begin{spacing}{1.0}
\maketitle

\begin{center}
{\it For John Cahn in honor of his seventieth birthday}
\end{center}
\end{spacing}

\begin{spacing}{1.0}
\begin{abstract}
The classical Lifshitz-Slyozov-Wagner 
 theory of domain coarsening predicts asymptotically 
self-similar behavior for the size distribution of a dilute system of particles
that evolve by diffusional mass transfer with a common mean field.
Here we consider the long-time behavior of measure-valued solutions
for systems in which particle size is uniformly bounded, 
i.e., for initial measures of compact support.

We prove that the long-time behavior of the size distribution 
depends sensitively on the initial distribution of the largest 
particles in the system.  Convergence to the classically predicted 
smooth similarity solution is impossible if the initial distribution function
is comparable to any finite power of distance to the end of the support.
We give a necessary criterion for convergence to other self-similar
solutions, and conditional stability theorems for some such solutions.
For a dense set of initial data, convergence to any self-similar solution
is impossible.
\end{abstract}

\footnotetext[1]{%
Institut f\"ur Angewandte Mathematik, Universit\"at Bonn, 
Wegelerstr.\ 6, 53115 Bonn, Germany}
\footnotetext[2]{%
Department of Mathematics \& 
Institute for Physical Science and Technology,
University of Maryland, College Park, MD 20742 USA}

\bigskip \null
\end{spacing}

\pagebreak
\begin{spacing}{1.0}


\section{Introduction}

The problem of Ostwald ripening is to explain the growth
of the typical particle size observed in the late-stage aging of
heterogeneously nucleated phase transitions.
In the classical theory of Lifshitz and Slyozov\cite{LS} and
Wagner,\cite{Wa} particles of the minority phase are assumed to 
be widely separated spheres that interact only through quasi-static 
diffusional exchange with a mean field. The chemical potential is proportional
to curvature on the particle boundaries, and its gradient provides
the flux that determines the particle growth rate.
For background to the problem and a review of the physical literature, 
see refs.~\citen{V,V2}.  

In the dilute approximation with appropriate units, 
the volume $v$ of any particle obeys the evolution law
\be
\frac{\D v}{\D t} = \vdot(v,\theta(t)) \equiv 
v^{1/3}\theta(t)-1,
\label{1vdot}
\ee
where $\theta(t)$ is the same for all particles.
The critical volume is $\theta(t)^{-3}=\frac{4\pi}{3}R_c(t)^3$,  
where $R_c(t)$ is the critical radius.
Conservation of mass determines the value of $R_c(t)$ or $\theta(t)$,
and if mass in the diffusion field can be neglected, total particle
volume is conserved and one finds that the critical radius equals
the average radius of currently existing particles. Particles with
radius larger that $R_c(t)$ are growing, while small particles  
shrink to zero size in finite time and disappear.

The LSW theory is concerned with a long-time regime in which most
initially existing particles have disappeared but a large number remain.
The system is characterized by the particle size distribution 
$f(t,v)$, which
we consider normalized so that $\int_0^v f(t,u)\,du$ is the number of
remaining particles with volume less than $v$, divided by the initial
number at time $0$. As particles of finite size are neither created 
nor destroyed, $f(t,v)$ should satisfy the conservation law
\be
\D_t f +\D_v(\vdot(v,\theta(t)) f)=0,
\label{1feqn}
\ee
where
\be
\theta(t) = {\int_0^\infty f(t,v)\,dv}
\left/{\int_0^\infty v^{1/3}f(t,v)\,dv}.\right.
\label{1theta}
\ee 
(These LSW equations also arise from a different model of particle clustering,
the Becker-D\"oring equations,
as recently shown by Penrose\cite{Pen}.)


Lifshitz and Slyozov\cite{LS} and Wagner\cite{Wa} argued that for large times,
the size distribution should behave in a self-similar manner.
They predicted: 
\begin{itemize}
\item[(a)] The critical radius cubed grows linearly for large time, 
like $4t/9$. 
\item[(b)] The particle size distribution approaches a form which
is self-similar under scaling by the critical radius; and
\item[(c)] In general, all scaled size distributions will approach the same
self-similar solution. This solution is smooth, compactly supported 
and explicitly computable. 
\end{itemize}

The basic structure of the scaling predictions (a) and (b) have been validated
in experiments, but the form predicted in (c) was not found. (Also, the particular
rate in (a) has proved difficult to measure.)
Breakdown of the law \eqref{1feqn}
(which implies that all particles of a given size have exactly the same
growth trajectory) 
is the most likely route for failure of the theory to explain experiment,
as has long been recognized.  
See ref.~\citen{N1} for a mathematical justification 
of this law by homogenization methods, that suggests that it is not 
necessarily valid, even in the limit of small volume fraction, if the 
electrostatic {\it capacity} of the particles does not also vanish.


In this paper we address the mathematical problem of whether
solutions of \eqref{1feqn} in fact must have the asymptotically 
self-similar form predicted by the LSW theory. 
In the LSW analysis it is clear that the equations permit a family of 
self-similar solutions that are not infinitely smooth at their point of vanishing. 
However, the arguments of Lifshitz and Slyozov suggest that these are unstable. 

The present results indicate that the long-time behavior of
the scaled particle size distribution depends sensitively on the behavior
of the initial distribution near the end of its support.
Let $\bar v(t)$ denote the supremum of the support of the volume
distribution; we assume $\bar v_0\colon=\bar v(0)<\infty$ throughout this paper.
We prove, for example, that if 
the initial distribution has the property that
\be
\psi_0(u):= \int_{\bar v_0-u}^{\bar v_0} f(0,v)\,dv\ge a u^p 
\qquad (u\ge0)
\label{1comp}
\ee
for  {\it some} exponent $p<\infty$ and some $a>0$, then
convergence to the smooth self-similar solution predicted by the LSW theory 
is impossible. 

The non-smooth self-similar solutions 
are characterized by their exponent of vanishing. 
Given any exponent $p$ in $(0,\infty)$ there is a unique (up to scaling) 
self-similar solution satisfying $\psi_0(u)\sim au^p$ as $u\to0$.
Our main result is to establish a condition on the initial data that is 
{\it necessary} for convergence to these self-similar distributions.
The condition is a little weaker than requiring that 
$\lim_{u\to0}\psi_0(u)/u^p$ exists
--- it states that the map $\psi_0$ must be ``regularly varying
at 0 with exponent $p$'' (see Definition \ref{D.regvar} and the books of Seneta\cite{Seneta}
and Bingham et al.\cite{Bing}).

We conjecture that this necessary condition is also {\it sufficient}
for convergence to a self-similar distribution characterized by a finite value of $p$.
Numerical simulations that we have performed tend to support this conjecture. 
We prove it for initial data sufficiently close in a certain sense to 
some self-similar solution corresponding to a sufficiently small value of $p$.
More generally, given arbitrary initial data that satisfy the necessary 
condition, we prove that the solution converges to self-similar form 
{\it if} the scaled inverse critical radius 
$\bar v(t)^{1/3}\theta(t)$  converges as $t\to\infty$.

By making an arbitrarily small change to any given initial distribution
in the vicinity of the tip of its support, one can easily arrange that
the convergence criterion holds, or does not hold, at will.  Our results
then imply that there is a dense set of initial data for which
convergence to {\it any} self-similar solution is impossible.  If it is
true that our necessary condition is also sufficient, then there is also
a dense set of data that yield convergence to any given self-similar
solution corresponding to a finite value of $p$.


The basic idea underlying these results is a simple one. That is the
notion that the essential effect of time evolution is to {\it stretch}
the size distribution, {\it chop off} the part corresponding to ``dead''
particles, and {\it normalize} to keep total volume constant.   In other
words, the action of the dynamics (up to a smooth distortion) is simply
to ``zoom in'' as with a microscope on the details of the particle size
distribution near the tip of its support.  

If the initial data behave in
an irregular fashion near this tip on a microscopic scale, we can expect
the solution to exhibit irregular behavior in time on a macroscopic
scale, as particles are extinguished in order of increasing size and the
distribution is normalized to conserve total volume. 
Indeed, the analysis in the proof of Theorem \ref{T.nec} below
shows that, with respect to the logarithm of distance to the tip of support,
the dynamics near the tip is rather well described by a continuous shift map.
This suggests that the dynamics determined by the 
LSW model \eqref{1feqn} may actually be {\it chaotic} in a mathematical sense
(with a dense trajectory and a dense set of time-periodic solutions).

It would be wrong to infer, however, 
that the dynamics of Ostwald ripening is chaotic
in any meaningful physical sense.
The mechanism that produces sensitive dependence of long-time behavior
on initial data in the LSW model depends heavily on the fact that
the growth rate is the same for all particles of the same size.
Lifshitz and Slyozov\cite{LS} already assumed some kind of breakdown 
in this rule in arguing that the non-smooth self-similar distributions would
not be observed physically due to the effect of ``encounters between different
grains.''  It is plausible that a theory that seeks to 
quantitatively explain the self-similar distributions 
observed in Ostwald ripening should 
account for dispersion of growth rates among same-size particles.


The notion that the asymptotic form of the size distribution
in LSW theory depends on the precise mode of vanishing of the 
initial distribution has emerged recently (independently of the present work) 
in work of Meerson \& Sasorov,\cite{MS} Giron et al.,\cite{GMS} and 
Carr and Penrose.\cite{CP}
The sole mathematical work in this connection is ref.~\citen{CP},
which concerns a simplified LSW theory for which the volume
growth rate is modeled by $v\theta(t)-1$, yielding linear characteristic
equations. Carr and Penrose prove convergence
to that self-similar solution characterized by $\psi_0(u)\sim au^p$
for initial data that have the property that 
$\int_0^x \psi_0(u)\,du = A(x)x^{p+1}$, where 
$p>0$ and $A$ is differentiable with $A(0)>0$. 
Furthermore, they give examples of initial data for which no convergence 
to self-similar form occurs,
and they also have results for data of infinite support.

Giron et al.\cite{GMS} expand a physical discussion given by
Meerson \& Sasorov\cite{MS}, and treat a class of models in which
$F(R,t)$, the number density of domains with radius $R$ at time $t$,
evolves according to a law of the form
\[
\frac{\D F}{\D t} +\frac{\D}{\D R}(VF)=0, \qquad
V(R,t)=\frac{1}{R^n}\left(\frac{1}{R_c(t)}-\frac{1}{R}\right)
\quad (n\ge -1).
\]
The critical radius $R_c(t)$ evolves so that the total volume
$\int_0^\infty R^3F(R,t)\,dR$ is conserved.
These authors assert that a size distribution initially describable by a 
power law $A_0(R_m(0)-R)^\lambda$ 
will for any $t>0$ have an expansion in which the leading
term has the form $A(t)(R_m(t)-R)^\lambda$, where the exponent $\lambda$ is
invariant in time. It is then asserted that if a self-similar 
asymptotic regime is ever reached, it must correspond to the self-similar
solution having the same $\lambda\in(0,\infty)$.
This assertion is supported by numerical computation
and a linearized asymptotic analysis at the tip of support, for initial data having the form 
$F_0(R)=\xi^\lambda g_1(\xi)+\xi^{\lambda_2} g_2(\xi)$ for $\xi=R_m(0)-R>0$, 
where $g_1$ and $g_2$ are analytic and $\lambda_2>\lambda>0$.
(The condition $\lambda>0$ corresponds to $p>1$ in the present paper.)

In the metallurgical literature, Brown (see refs.~\citen{B1,B2,B3} 
and the references therein) had earlier asserted that an arbitrary initial
distribution evolves to that self-similar solution ``with the same tail,''
but by this he appears to always mean infinite-support self-similar
solutions, for which the volume $\int_0^\infty v f(t,v)\,dv$ is 
infinite.  Regarding distributions of compact support (which he
calls ``discontinuous''), he asserts they evolve to the distribution
predicted by LSW. E.g., in ref.~\citen{B2} he says ``Both Hillert 
et al.\ (4) and Brown (5) have shown that an initial triangular distribution
develops into the LSW distribution, albeit slowly.''
This kind of assertion is contradicted by the results we prove in this paper.
However, Brown does make a number of persuasive arguments 
regarding the physical meaning of perturbing the distribution in the 
range of the largest particles in the system, and the ultimate dominance
of the distribution of the largest particles. 
For example, in ref.~\citen{B3} he mentions
that the {\it rate} of coarsening can be slowed if the distribution
favors the largest particles (as it does for self-similar
solutions corresponding to small values of the exponent $p$).


The structure of this paper is as follows. In Section 2 we reformulate
the evolution law for the size distribution and recall the results
proved in ref.~\citen{NP1} concerning the well posedness of the initial value
problem. We show that the solution is given for finite time by an analytic
distortion of the initial data. Section 3 contains results for long time
that involve the unscaled variables. In particular, if the initial
size distribution contains a Dirac delta at the tip of the support,
the distribution converges to a stationary Dirac delta distribution
as $t\to\infty$, and if not, then $\bar v(t)$ increases to infinity.
In Section 4, we classify self-similar solutions by exponent of vanishing.
In Section 5 we scale the distribution function 
by the maximum particle volume $\bar v(t)$ 
and establish a variety of results as indicated above, concerning
long-time convergence in the rescaled variables.
We conclude in Section 6 with a discussion, and give examples of initial
data yielding convergence and nonconvergence to self-similar form.

\section{Well posedness, regularity, stretching}

\subsection{A well-posed formulation}

Some care should be taken in the physical interpretation of 
stability and instability results asserted for particle size distributions. 
One ought to consider what kinds of particle volume distributions are 
physically meaningful, and what kinds of perturbations make sense.
It is plausible that in late-stage coarsening it is not possible to 
nucleate particles that are indefinitely large, for example.
And it seems reasonable to imagine that small perturbations to the system 
would involve small changes to particle volumes.

In this article we treat solutions for which the initial size
distribution $f(0,v)\,dv$ is an arbitrary probability measure with
compact support. This admits distributions for which a finite
fraction of the particles all have the same size, for example.
Compact support corresponds to the requirement that particle sizes 
initially are bounded. 

A theory for well posedness of the initial value problem for such
solutions is developed in ref.~\citen{NP1}. 
(For a family of closely related Lifshitz-Slyozov models, 
Collet and Goudon\cite{CG} have recently studied the initial-value problem 
for initial distributions $f(0,\cdot)$ that are integrable with finite 
first moment.) 
To discuss the results of ref.~\citen{NP1}, it is
better to work with an equivalent formulation using different variables.
First, let $\F(t,v)$ denote the fraction of initially existing particles
with volume $\ge v$ at time $t$. 
With $F(v)=1-\F(t,v)$, $F$ is the usual distribution function for the
particle volume distribution. 
When there is no danger of confusion we also refer
to $\F(t,\cdot)$ as the ``distribution function $\F$.'' 
The function $v\mapsto\F(t,v)$ is decreasing and left-continuous
at jumps with $\F(t,0)=1$.
(To be precise, we say $\F$ is {\it decreasing} if $\F(x_1)\le \F(x_2)$
whenever $x_1\ge x_2$, and similarly with {\it increasing}. A decreasing
function need not be strictly decreasing.)
The particle volume distribution is a measure, 
in general, given formally by $f(t,v)\,dv=dF(v)$,
that is, $\F(t,x)=\int_x^\infty f(t,v)\,dv$.  

Next we invert the relationship between $v$ and $\F$ and
consider the volume $v$ at time $t$ as a
function of $\F\in[0,1]$, taken so that $\F\mapsto v(t,\F)$
is decreasing and right-continuous at jumps with $v(t,1)=0$.
For a finite system of particles with volumes ranked 
in decreasing order $v_0(t)\ge\ldots\ge v_{N-1}(t)$, we have
$v(t,\F)=v_j$ for $\F\in[\frac jN,\frac{j+1}N)$.   
In general, 
\be \label{2vdef}
v(t,x)=\sup\{y\mid \F(t,y)>x\} \quad\mbox{for $0\le x<1=\max\F$.}
\ee
We shall call the map $\F\mapsto v(t,\F)$ the {\it volume ranking}
for the system at time $t$, and refer to $\F$ as the {\it particle rank}.
The volume ranking determines the distribution function according
to the prescription
\be
\F(t,y)=
\sup\{x\mid v(t,x)>y\}\vee 0 =
\cases{
\sup\{x\mid v(t,x)>y\} & for $0\le y<\max v(t,x)$, \cr
0 & for $y\ge\max v(t,x)$.}
\label{2inv}
\ee

The notion of distance between size distributions employed in the
well-posedness theory of ref.~\citen{NP1}  corresponds to the
supremum-norm distance between the associated volume rankings.
This is the least ``maximal volume change'' required to alter
one size distribution into the other. Mathematically this corresponds to
the notion of the $L^\infty$ Wasserstein distance between probability
measures, as shown in ref.~\citen{NP1}.

The equation governing the evolution of the volume ranking
$v(t,\F)$ is just \eqref{1vdot}.
In ref.~\citen{NP1}, we proved the following theorem on the global 
well-posedness of the initial value problem.
Let $\rcd$ be the set of right-continuous decreasing 
functions $v_0\colon[0,1]\to\R$ with $v_0(1)=0$. 
On $\rcd$ we have the metric topology 
given by the supremum norm of the difference:  
\[
\|v_1-v_2\|=\sup_\F|v_1(\F)-v_2(\F)|.
\]
For any $T>0$, $C([0,T],\rcd)$ is the metric space of continuous 
$v\colon[0,T]\to\rcd$,
with metric given by $\sup_{[0,T]}\|v_1(t,\cdot)-v_2(t,\cdot)\|$.

\begin{theorem} \label{T.ivp}
Given $v_0\in\rcd$, there is a unique continuous
$v\colon[0,\infty)\to\rcd$ and a unique $\theta\in L^\infty_\loc(0,\infty)$
such that 
\[
\int_0^1 v(t,\F)\,d\F = \int_0^1 v_0(\F)\,d\F
\]
for all $t\ge0$, and
\[
v(t,\F)= v_0(\F)+\int_0^t (v(s,\F)^\third\theta(s)-1)\,ds
\]
for all $(t,\F)$ such that $v(t,\F)>0$.
Moreover, the map from $v_0$ to $v$ is locally Lipschitz continuous from
$\rcd$ into $C([0,T],\rcd)$, for any $T>0$.
\end{theorem}

The value of $\theta(t)$ is determined by conservation of volume to be
\be
\theta(t) = \bar \F(t)\left/ \int_0^{\bar\F(t)} v(t,\F)^{1/3}\,d\F \right. 
\label{2theta}
\ee
where $\bar \F(t)$ is the end of the support of $v(t,\cdot)$, i.e.,
\be
\bar\F(t)=\sup\{\phi\in[0,1]\mid v(t,\phi)>0\}.
\label{2Fbar}
\ee
The function $\bar\F$ is decreasing in time, but may have jumps.
Consequently $\theta$ can jump, but it is positive and bounded on finite
intervals of time.  For fixed $\F$, $t\mapsto v(t,\F)$ is Lipschitz
continuous, and satisfies \eqref{1vdot} for almost every $t$,
as long as $v(t,\F)>0$. 

\subsection{Stretching}

The result below indicates that for finite time the action
of the dynamics {\it stretches} the volume ranking vertically in a smooth
manner (by an analytic map). Thus, we can anticipate that whatever
limiting behavior or irregularity is present in the initial data $v_0(\F)$
in the limit $\F\to0$ will be preserved for any finite length of time.

\begin{prop} \label{P.analytic}
Let $(\theta,v)$ be a solution of \eqref{1vdot} as given by 
Theorem \ref{T.ivp}.
For $x>0$, let $\chv(t,x)$ be the solution of 
\[
\chv(t,x)=x+\int_0^t (\chv(s,x)^{1/3}\theta(s)-1)\,ds
\]
defined on the maximal time interval $[0,\hat t(x))$ where $\chv>0$. Then
\begin{itemize}
\item[(a)] $\chv$ is analytic in $x$, i.e., whenever $\chv(t_0,x_0)>0$, the map
$x\to \chv(t_0,x)$ is analytic near $x_0$.
\item[(b)] $\D \chv/\D x$ is strictly increasing in time for each $x$.
\item[(c)] $ v(t,\F) = \chv(t,v_0(\F)).  $
\end{itemize}
\end{prop}

\begin{proof}
The formula in (c) follows by the uniqueness of solutions to the 
initial value problem for \eqref{1vdot}.

Since the function $w\mapsto w^{1/3}$ is smooth for $w>0$,
a standard proof shows that $x\mapsto \chv(t_0,x)$ is continuously differentiable
to any order, and each derivative is Lipschitz continuous in time. 
Almost everywhere in time, the derivative satisfies
\[
\frac{\D}{\D t}\frac{\D \chv}{\D x}(t,x) =  
\frac{\theta(t)}{3\chv(t,x)^{2/3}} 
\frac{\D \chv}{\D x}(t,x), \quad 
\frac{\D \chv}{\D x}(0,x) = 1 ,
\]
therefore
\[
\frac{\D \chv}{\D x}(t,x) = \exp \int_0^t
\frac{\theta(s)}{3\chv(s,x)^{2/3}}  \,ds,
\]
and this is strictly increasing in time. This proves (b).

To prove (a) we use a result related to a theorem of Bernstein.
We recall that a smooth function
$g\colon(0,\infty)\to\R$ is called {\it completely monotone} if 
its derivatives satisfy $ (-1)^k g^{(k)} \ge 0$ for all integers
$k\ge0$. As a consequence of Bernstein's theorem that every completely
monotone function has a representation as the Laplace transform of
a positive measure, it follows that every completely monotone function
has an analytic extension to the open right half of the complex plane.
(See ref.~\citen{BF}, for example, for a proof.) 

We note that if $\chv(t_0,x_0)>0$, then it is easy to show
that $\chv(t_0,x_0)<\chv(t_0,x)<\infty$ for all $x>x_0$.
Thus, to prove (a) it suffices to show that if $\chv(t_0,x_0)>0$,
then the function $x\mapsto(\D \chv/\D x)(t_0,x_0+x)$ is completely
monotone. To do this, we show that the function given by 
\[
\omega(t,x)= -\chv(t,-x+x_0) \quad\mbox{for $0\le t\le t_0$, $x<0$,}
\]
satisfies $\D_x^k\omega\ge0$ for all integers $k\ge1$.
This is true for $k=1$ by (b). Suppose it holds for $1\le k<j$ for some $j$.
Let $G(y)=-(-y)^{1/3}$ for $y<0$, and note that the derivatives $G^{(k)}$ are
positive for all $k\ge1$. The function $\D_x^j\omega$ satisfies
\[
\frac{\D}{\D t} \D_x^j\omega = \theta(t)\D_x^j(G(\omega)) = 
\theta(t)(G'(\omega)\D_x^j\omega+R_j(t,x)),
\]
where $R_j(t,x)$ is a sum of terms of the form
$c_\alpha G^{(k)}(\omega)
\D_x^{\alpha_1}\omega\ldots\D_x^{\alpha_n}\omega$
which have positive coefficients and involve derivatives
$\D_x^k\omega$ of order $\le j-1$. Then it follows that 
$R_j(t,x)\ge0$ and hence
\[
\D_x^j\omega(t,x) = \int_0^t\exp\left(\int_s^t G'(\omega(r,x))\,dr\right)
\theta(s) R_j(s,x)\,ds \ge 0
\]
for $0\le t\le t_0$, $x<0$. We obtain the desired result by induction.
\eproof
\end{proof}

\subsection{Evolution of the distribution function}

For the main part of our study of long-time behavior we will
work with the distribution function $\F(t,v)$, rescaled appropriately to 
study the development of self-similar form.
To determine the time evolution of $\F(t,v)$, we need to use \eqref{2inv} to 
get the appropriate generalized inverse function.
The following result describes the evolution and
implies that it is a solution of the advection equation
\be
\D_t\F +\vdot(v,\theta(t))\D_v\F = 0
\label{2Feqn}
\ee
whenever the initial data $\F_0=\F(0,\cdot)$ is differentiable.
The characteristics of \eqref{2Feqn}  are given by the function 
$\chv$ of Proposition \ref{P.analytic}.

\begin{lemma} \label{L.Fevol}
Let $\F(t,y) = \sup\{ x\mid v(t,x)>y\} \vee0$.
Then 
\[
\F(t,\chv(t,y))= \F_0(y)
\]
whenever $0<\chv(t,y)<\bar v(t)$.
\end{lemma}

\begin{proof}
Using Proposition \ref{P.analytic} we find that
\[
\F(t,\chv(t,y))=
\sup\{x\mid \chv(t,v_0(x))>\chv(t,y)\} =
\sup\{x\mid v_0(x)>y\} = \F_0(y),
\]
provided $0<\chv(t,y)<\bar v(t)$.
\eproof
\end{proof}
\section{Long-time behavior I }

In this section we derive some basic results as time tends to infinity 
without scaling the solution.
Below we shall denote the maximum particle volume at time $t$ by 
\[
\bar v(t) = v(t,0)=\sup_\F v(t,\F).
\]
Note that from \eqref{2theta} it follows 
\[
\bar v(t)^\third\theta(t)\ge1,
\]
therefore $\bar v(t)$ is {\it increasing}.
The total volume of particles in the system will be written
\[
V = \int_0^1 v(t,\F)\,d\F.
\]

\subsection{The case of a Dirac delta at the tip}

First we consider the case in which a positive fraction of the particles 
have volume equal to the maximum possible in the system.
In terms of the particle volume distribution $f$ this means we consider
measure-valued initial data $f_0$ that carry a Dirac delta at the tip of the support. 
That is, 
\[
f_0(v)=a \delta{(v-\bar v_0)}+ \tilde f_0(v)
\]
where $0<a\le1$, $\delta{(v-\bar v_0)}$ denotes the Dirac delta distribution at $\bar v_0$
and 
\[
\lim_{v \to \bar v_0} \int_v^{\infty} \tilde f_0(y)\, dy =0.
\]
For the volume ranking function $v$  this assumption means that
the solution $v(t,\cdot)$ is constant on the interval $[0,a)$.
In this case we can characterize the asymptotic behavior as follows.

\begin{prop} \label{P.atom}
Assume that for some $a\in(0,1]$, the initial data satisfy
$v_0(0)=v_0(\F)$ for $0\le\F<a$ and $v_0(0)>v_0(\F)$ for $\F>a$. 
Then
\[
\lim_{t\to\infty}\bar\F(t)=a,
\]
and we have
\[
\lim_{t\to\infty} v(t,\F)=\cases{ V/a, & $0\le\F<a$, \cr 0,& $\F>a$.}
\]
Correspondingly,
\[
\lim_{t\to\infty} \F(t,v)= \cases{a, & $0\le v<V/a$, \cr 0,& $v>V/a$.}
\]
\end{prop}

\begin{proof}
From the assumption on the initial data it follows that for all $t$, 
$v(t,\F)=\bar v(t)$ for $0\le\F<a$ and $v(t,\F)<\bar v(t)$ for $\F>a$.
From conservation of volume it follows
\[
V=\int_0^{\infty} v (t,\F)\,d\F \geq a \bar v(t).
\]
Hence $\bar v(t)$ is uniformly bounded, and since it is 
increasing the limit $v_\infty=\lim_{t\to\infty}\bar v(t)$ exists.

Assume now that 
\[
\lim_{t \to \infty} \bar \F(t) > a.
\]
(Since $\bar v(t)\ge V>0$, $\bar \F(t)\ge a$ for all $t$.)
Then there exists $\F >a$ such that $v(t,\F) >0$ for all $t>0$.
But then for almost every $t>0$, using the inequality
$a-b\ge(a^3-b^3)/3a^2$ for $a>b>0$ we have 
\beqna
\D_t (\bar v (t) - v(t,\F)) &=& \theta(t) \left(\bar v (t)^\third - v(t,\F)^\third\right)
\\&\geq& \frac{\theta(t)\bar v(t)^\third}{3\bar v(t)} (\bar v(t)-v(t,\F))
\\&\geq& \frac{a}{3V} (\bar v(t)-v(t,\F)),
\eqna
and hence
\[
\bar v (t) - v(t,\F) \geq ( \bar v(0) - v_0(\F)) e^{at/3V}.
\]
Since by assumption $\bar v(0) - v_0(\F)>0$ and since $\bar v(t)$ is bounded
this contradicts the presumption that $v(t,\F)>0$ for all $t$.
It follows that for each $\F>a$, $v(t,\F)=0$ for sufficiently large $t$,
therefore $\lim_{t \to \infty} \bar \F(t) =a$. 

That $v_\infty=\lim_{t\to\infty}\bar v(t)=V/a$ follows by volume conservation.
The limiting behavior asserted for $\F(t,v)$ follows from the fact that
$a\le\F(t,v)\le\bar\F(t)$ for $0\le v<\bar v(t)$ and $\F(t,v)=0$ for 
$v>\bar v(t)$.
\eproof
\end{proof}

\subsection{The case of no Dirac delta at the tip}
Here and throughout the rest of this work we will assume that no positive fraction
of the particles have the maximal particle volume $\bar v$. 
For the initial
volume distribution $f_0$ this means
\[
\lim_{v \to \bar v_0} \int_{v}^{\infty} f_0(y)\,dy =0.
\]

\begin{prop} \label{P.noatom}
Assume that $v_0(0)>v_0(\F)$ for all $\F\in(0,1]$. Then 
\begin{itemize}
\item[(a)] $\lim_{t\to\infty} \bar v(t)=\infty$, i.e., the maximal particle volume tends to 
infinity.
\item[(b)] $\liminf_{ t \to \infty} \theta(t) =0,$ meaning the critical radius is unbounded.
\item[(c)] $\lim_{t\to\infty} \bar\F(t)=0$, i.e., the fraction of initially existing particles 
that continue to exist at time $t$ decreases to zero.
\end{itemize}
\end{prop}
We do not know whether the result in (b) can be improved 
to say $\lim_{t\to\infty}\theta(t)=0$. 

\begin{proof}
We prove that $\bar v(t)$ is unbounded using volume conservation.
Suppose that $\bar v(t)\le C$ for all $t$. As in the proof of Proposition \ref{P.atom}
we find that for any $\F>0$, as long as $v(t,\F)>0$ we have
\[
\bar v (t) - v(t,\F) \geq ( \bar v(0) - v_0(\F)) e^{t/3C}.
\]
Consequently, for each $\F$, $v(t,\F)=0$ for sufficiently large $t$.
This means that $\bar\F(t)\to0$ as $t\to\infty$. Then we infer that
$V \le \bar v(t)\bar\F(t) \to0$, so $V=0$, but the hypotheses guarantee $V>0$.
Hence $\bar v(t)$ is unbounded; since it is increasing, it tends to infinity.

Suppose the claim in (b) fails. Then since $\theta(t)$ is bounded away from zero
for $t$ in any finite interval, there is some $\delta>0$ such that 
\[
\theta(t)\ge\delta
\]
for all $t$. From (a) and the fact that 
$\F\mapsto v(t,\F)$ is right continuous, there exists $\eps>0$ and 
$(t_0,\F_0)$ with $\F_0>0$ such that $v(t_0,\F_0)^\third\delta-1>\eps$, 
which implies that 
\[
\frac{\D v}{\D t}(t,\F_0) = v^\third\theta(t)-1 \ge \eps >0
\]
near $t=t_0$ and so for almost every $t\ge t_0$. This yields that $v(t,\F_0)$ is unbounded,
and this contradicts the inequality
\be
V = \int_0^1 v(t,\F)\,d\F \ge \F_0 v(t,\F_0),
\label{3vbdd}
\ee
which is valid for all $t$ since $\F\mapsto v(t,\F)$ is decreasing. This proves (b).

To prove (c), note that since $v^\third<1+v$ we have
\[
\bar \F(t) = \theta(t) \int_0^1 v^{\third}(t,\F) \,d\F \le \theta(t)(1+V).
\]
By (b) there exists a sequence $t_n \to \infty$
such that $\theta(t_n) \to 0$. Since $\bar \F(t) $ is decreasing the
assertion (c) follows.
\eproof
\end{proof}

\section{Self-similar solutions}

For understanding the long-time dynamics, we shall study the
distribution function or particle rank $\F$ as a function of volume
$v$ according to the dynamical formulation described
in Lemma \ref{L.Fevol}. (Recall that
the usual distribution function $F(t,v)=1-\F(t,v)$.)

We look for solutions in self-similar form.
We begin with some general considerations, temporarily 
allowing for the possibility that the volume distribution has infinite support. 
Any self-similar distribution function $\F$ for which total particle volume is 
finite and conserved must have the form
\be
\F(t,v)= a(t)\preSF(a(t)v)
\label{S.form2}
\ee
for some decreasing functions $\preSF$ and $a$. 
The total volume satisfies
\[
V = \int_0^\infty \F(t,v)\,dv = \int_0^\infty \preSF(s)\,ds,
\]
and we have
\be
\theta(t) = \frac{\F(t,0)}{ \int_0^\infty 
\frac13 v^{-2/3}\F(t,v)\,dv}
= \frac{a(t)^\third \preSF(0)}{ \int_0^\infty
\frac13 s^{-2/3}\preSF(s)\,ds.}
\label{Sthet}
\ee
\begin{lemma} \label{L.cmpct}
Any self-similar distribution function $\F$ in the form \eqref{S.form2}
with finite total volume has compact support.
\end{lemma}

\begin{proof}
Given the form in \eqref{S.form2}, by \eqref{Sthet} we may scale 
$a$ and $\preSF$ so that $\theta(t)a(t)^{-1/3}\equiv1$.
According to Lemma \ref{L.Fevol}, being a solution requires that
\[
 a(t)\preSF(a(t)\chv(t,y)) = a(0)\preSF(a(0)y)
\]
whenever $0<\chv(t,y)<\bar v(t)$.
For a nonconstant solution,  $a$ and $\preSF$ should be 
Lipschitz, and differentiating we find that 
\[
0= a'(t)(\preSF+s\preSF'(s))+a^2 (s^\third-1)\preSF'(s).
\]
Separating variables, we find that 
$a'/a^2$ must be a negative constant.
We set $b=-a^2/a'$, then $a(t)=ba(0)/(b+a(0)t)$,
and we must have
\be
0 = \preSF(s)+(s+b(1-s^\third))\preSF'(s).
\label{S.Heq1}
\ee
As long as $\preSF>0$ we compute that 
\be
-\ln \left(\frac{\preSF(s)}{\preSF(0)}\right)
 = \int_0^s \frac{dx}{x+b(1-x^\third)} .
\label{S.Heq2}
\ee
The function $s\mapsto s+b(1-s^\third)$ is convex with the 
value $b>0$ at $s=0$, and at $s=(b/3)^{3/2}$ it achieves its
minimum value $b(1-\sqrt{4b/27})$. This is positive if 
$b<27/4$ and nonpositive if $b\ge27/4$. 

In the former case, we have $\preSF(s)>0$ for all $s>0$ and we
compute that
\[
\lim_{s\to\infty} -\ln \left( \frac{\preSF(s)}{\preSF(0)}
\left(\frac{s+b}b\right)\right)
=\int_0^\infty \left(\frac1{x+b(1-x^\third)}-\frac1{x+b}\right)dx
\]
and this is finite and positive. Hence $\preSF(s)\sim C/s$ as $s\to\infty$
for some $C>0$, and this means that the volume 
$\int_0^\infty \preSF(s)\,ds$ diverges.

In the case $b\ge27/4$, on the other hand, the integral in \eqref{S.Heq2}
diverges as $s\to s_0$, where $s_0$ is the smallest positive zero
of $s+b(1-s^\third)$. 
(This zero is between $1$ and the value $(27/4)^{3/2}/3\sqrt3=27/8$,
and decreases to $1$ as $b\to\infty$.)
Then $\preSF(s)\to0$ as $s\to s_0$, and we must put
$\preSF(s)=0$ for $s\ge s_0$.
\eproof
\end{proof}

Thus we see that all self-similar solutions with finite total volume
have compact support. To describe these solutions in a way that
connects with the analysis to come,
we rescale with respect to the tip of the support, introducing
the variables $u$ and $\SF(u)$ that satisfy
\be
1-u =\frac{v}{\bar v(t)}= \frac{s}{s_0},
\quad \SF(u)= \frac{\bar v(t)\F(t,v)}{V} = \frac{s_0\preSF(s)}{V}.
\label{Svars}
\ee
Note $v=s/a(t)$ and $\bar v(t)=s_0/a(t)=\bar v(0)+ts_0/b$.
We set
\[
\ks = \frac{s_0+b}{3s_0}=\frac{bs_0^\third}{3s_0},
\]
then $b/s_0=3\ks-1$ and $s_0=(3\ks/(3\ks-1))^3$, and
$\ks$ increases from 1 to $\infty$ as $b$ increases from
$27/4$ to $\infty$. Furthermore we have
\[ 
\frac{s+b(1-s^\third)}{s_0} = u(\ks Q(u)-1),
\]
where
\be
Q(u) = 3\left(\frac{1-(1-u)^\third}u\right) .
\label{SQdef}
\ee
The function $Q$ satisfies $Q(0)=1$, $Q(1)=3$, 
and is given by the power series 
\[
Q(u)=1+\frac13 u + \frac5{27}u^2+\ldots,
\]
which converges for $|u|<1$ and whose coefficients are all positive.

The self-similar profiles $\SF(u)$ are increasing functions that 
satisfy
\be
\frac{d}{du} \ln\SF = \frac1{u(\ks Q(u)-1)} 
\label{SSFeq}
\ee
for $0<u<1$, with the normalization
\be \label{SSvol}
\int_0^1\SF(u)\,du=1.
\ee
Note that for $\ks>1$,
\[
\frac1{u(\ks Q(u)-1)}
= \frac1{(\ks-1)u}-\left(\frac{\ks}{\ks-1}\right)
\frac{Q(u)-1}{u(\ks Q(u)-1)}.
\]
The last term is negative and remains bounded as $u\to0$.
Integrating \eqref{SSFeq}, we find that self-similar solutions
may be characterized according to the exponent $p=1/(\ks-1)$,
which governs the rate of vanishing as $u\to0$. 
Henceforth, we write $\SFP$ to denote the solution $\SF$ of 
\eqref{SSFeq}--\eqref{SSvol} with $\ks=1+1/p$.

\begin{lemma}\label{L.sim}
For any $p\in(0,\infty]$, equation \eqref{2Feqn}
for the distribution function or particle rank has a self-similar solution
determined by \eqref{SSFeq} and \eqref{Svars} with $\ks=1+p^{-1}$.
For $p<\infty$ the profile has the form
\[
\SFP(u) = \alpha_p(u) u^p
\]
where $\alpha_p$ is decreasing
and analytic on $[0,1)$.
For $p=\infty$ ($\ks=1$) we have $\SFI(u)=o(u^q)$ as $u\to0$ for all $q>0$.
\end{lemma}

Equation \eqref{SSFeq} can be explicitly integrated using the substitution
$1-u=x^3$ and the expansion
\[
u(\ks Q(u)-1)= (x-1)(x-a)(x+a+1)
\]
where $2a=-1+\sqrt{3(4\ks-1)}$.  For $p<\infty$ ($\ks>1$) we find
\be
\frac{\SFP(u)}{\SFP(1)} = \frac{(1-x)^{p}}
{ \left(1-\frac xa\right)^{p_1} \left(1+\frac x{a+1}\right)^{p_2}}
\label{SFPeq}
\ee
where we have
\[
p= \frac{3}{(a-1)(a+2)} ,\quad
p_1= \frac{3a^2}{(a-1)(2a+1)} ,\quad
p_2= \frac{3(a+1)^2}{(a+2)(2a+1)} .
\]
In ref.\ \citen{GMS} analogous expressions are given for the corresponding
particle distribution $\Phi(\tau,x):= \partial_u \Psi(\tau,u)$ where
$1-u=x^3$.
The solution singled out by Lifshitz, Slyozov and Wagner
is that corresponding to $p=\infty$
or $\ks=1$ and satisfies
\be
\frac{\SFI(u)}{\SFI(1)} = \frac{ e^{-x/(1-x)}}
{ (1-x)^{5/3}(1+x/2)^{4/3}}.
\ee
When interpreting these formulae, note that
\[
x = (1-u)^\third = \left( \frac{v}{\bar v(t)}\right)^\third =
\frac R{\bar R(t)}
\]
where $R$ represents particle radius and $\bar R(t)$ is the maximal
particle radius.

As $p\to0$ in \eqref{SFPeq} we have $a\to\infty$ and we see that 
$\SFP(u)\to1$ for  $0\le u<1$. That is, the particle distribution concentrates 
into a Dirac delta at the tip of the support. 
For the self-similar solution with profile $\SFP$, the maximal particle volume
satisfies  
\[
\bar v(t)=\bar v(0)+ \frac{pt}{2p+3}.
\] 
Thus, as Brown\cite{B3} indicated, 
the rate of coarsening slows to zero as the particle distribution concentrates
at the tip, which means $p\to0$.

\section{Long-time behavior II }

\subsection{Rescaled variables}
In the following we always consider the case that no positive
fraction of particles has the maximal volume $\bar v$.
That is, we assume that the volume distribution
carries no atom at the tip. To study the emergence of
self-similar behavior for large times, it is appropriate
to scale the volume by the maximal volume $\bar v(t)$.
We introduce new variables via
\be
\tau = \ln \frac{\bar v(t)}{\bar v(0)}, \quad
u= 1-\frac{v}{\bar v(t)}, \quad
\psi= \frac{\bar v(t)\F}{V}. 
\label{IIvar}
\ee
For each $\tau\ge0$, the function $u\mapsto\psi(\tau,u)$ is
left continuous and increasing on $[0,1]$ with $\psi(\tau,0)=0$.
With $\psi_0(u)=\bar v(0)\F_0(v)/V$, Lemma \ref{L.Fevol} implies that the evolution
of $\psi(\tau,u)$ is determined by the relation
\be
\psi(\tau,\chu(\tau,u))= e^\tau\psi_0(u),
\label{IIevol1}
\ee
as long as $0<\chu(\tau,u)<1$, 
where $\chu(\tau,u)=1-\chv(t,v)/\bar v(t)$ satisfies
\be
\frac{\D\chu}{\D \tau} =  (\ka(\tau) Q(\chu)-1)\chu, \quad
\chu(0,u)=u.
\label{IIchar1}
\ee
Here $Q$ is the function defined in \eqref{SQdef}, 
and $\ka(\tau)>\frac13$ is determined from
\be
\frac{1}{\bar v(t)^\third\theta(t)} = 1-\frac1{3\ka(\tau)}=
\frac1{\psi(\tau,1)} \int_0^1\frac13 (1-u)^{-2/3} \psi(\tau,u)\,du.
\label{IIka}
\ee
The conservation of total volume is expressed by the identity
\be
 \int_0^1 \psi(\tau,u)\,du = 1,
\label{IIvol}
\ee
valid for all $\tau\ge0$.

For differentiable initial data, $\psi$ satisfies 
\be
\D_\tau\psi + (\ka(\tau)Q(u)-1)u\D_u\psi = \psi
\label{IIevol2}
\ee
for $0<u<1$, $\tau>0$. The characteristics of this PDE satisfy
\eqref{IIchar1}. Note that $\ka(\tau)Q(1)-1>0$ for all $\tau$,
and therefore every point $(\tau_1,u_1)\in(0,\infty)\times(0,1)$
lies on a unique characteristic that can be continued back to time
$\tau=0$. That is, $u_1=\chu(\tau_1,\tilde u_1)$ for some
$\tilde u_1\in(0,1)$.

Self-similar solutions of \eqref{2Feqn} correspond to 
stationary solutions of \eqref{IIevol2}, for which we know
$\ka=\ks\ge1$ is constant and $\psi=\SFP$ where $p=1/(\ks-1)$.
In general we are interested in investigating under what conditions
$\psi(\tau,u)$ converges as $\tau\to\infty$. 
We shall see that the long-time behavior of solutions is a sensitive
function of the manner in which the initial data vanish as $u\to0$.

\subsection{Tip behavior and the mean field}

\begin{prop}\label{P.comp}
Let $\psi$ be determined by \eqref{IIevol1}--\eqref{IIvol},
and suppose $0<p<\infty$.
\begin{itemize}
\item[(a)] If $\ \inf_{u>0}{\psi_0(u)}/{u^p}>0$, then
$\ \limsup_{\tau\to\infty} \ka(\tau)\ge 1+\frac1p $.
\item[(b)] If $\ \sup_{u>0}{\psi_0(u)}/{u^p}<\infty$, then
$\ \liminf_{\tau\to\infty} \ka(\tau)\le 1+\frac1p $.
\end{itemize}
\end{prop}

\begin{proof}
By Lemma \ref{L.sim}, given the hypotheses of part (a), 
there exists $a_0>0$ such that $\psi_0(u)\ge a_0\SFP(u)$ for $0<u<1$.
We aim to construct a ``subsolution'' of the form 
\[
\psi_-(\tau,u)= a(\tau)\SFP(u)
\]
by taking $a(0)=a_0$ and ensuring that $\psi/\psi_-$ 
is increasing along the characteristics given by \eqref{IIchar1}. 
By \eqref{IIevol1} it is equivalent to show the map 
$\tau\mapsto \tau-\ln\psi_-(\tau,\chu(\tau,u))$
is increasing. Using \eqref{SSFeq}, we find that it suffices to ensure that 
$a(\cdot)$ is Lipschitz and for almost every $\tau>0$ satisfies
\be
\frac{a'(\tau)}{a(\tau)}
\le 1- \frac{\ka(\tau)Q(u)-1}{\ks Q(u)-1}
= \frac{\ks-\ka(\tau)}{\ks-1/Q(u)}
\label{IIa1}
\ee
for $0<u<1$. Since $1\le Q(u)\le3$ this is guaranteed if we require
\begin{eqnarray}
\frac{a'(\tau)}{a(\tau)} 
&=& 
\cases{
(\ks-\ka(\tau))/(\ks-1), & if $\ks-\ka(\tau)>0$,\cr
(\ks-\ka(\tau))/(\ks-\frac13), & if $\ks-\ka(\tau)<0$,\cr
0,& if $\ks-\ka(\tau)=0$. }
\label{IIa2}
\end{eqnarray}
With this choice, it follows $\psi(\tau,u)/a(\tau)\SFP(u)\ge1$
for $0\le u \le1$ and for all $\tau\ge0$. 

Suppose now that $\limsup\ka(\tau)<\ks$. Then $\liminf a'(\tau)/a(\tau)>0$
and it follows $a(\tau)\to\infty$ as $\tau\to\infty$. But 
this contradicts volume conservation. Hence $\limsup\ka(\tau)\ge\ks$.

Part (b) is proved in similar fashion. One constructs a supersolution
of the form $\psi_+(\tau,u)=a(\tau)\SFP(u)$, ensuring that
the inequality in \eqref{IIa1} is reversed by interchanging the
first two cases of \eqref{IIa2}.
If $\liminf\ka(\tau)>\ks$ then $a(\tau)\to0$ and again using volume 
conservation one obtains a contradiction.
\eproof
\end{proof}

\subsection{Necessary conditions for convergence}

Our aim in this section is to derive restrictions
on initial data $\psi_0$ that must hold if convergence occurs
in the rescaled variables.

\subsubsection{Consequences of convergence} 

\begin{lemma} \label{L.lim}
Assume that $\lim_{\tau\to\infty}\psi(\tau,u)=\psii(u)$ exists
for each $u\in[0,1]$. Then $\lim_{\tau\to\infty}1/\ka(\tau)=1/\kai$
exists, where $1\le\kai\le\infty$, and $\psii(u)=\SFP(u)$, where 
$p=1/(\kai-1)\in[0,\infty]$.
Moreover, as $\tau\to\infty$, $\psi(\tau,u)$ converges
uniformly for $u$ in any compact subset of $(0,1]$.
\end{lemma}

\begin{proof}
Since $\psi(\tau,u)$ is increasing in $u$, the dominated convergence
theorem implies that $\lim_{\tau\to\infty}1/\ka(\tau)$ exists. If the limit
is zero, then since $\psii$ is increasing it follows from \eqref{IIka}--\eqref{IIvol}
that $\psii(u)=\psii(1)=1$ for all $u\in(0,1]$. This is the case $p=0$ of the Lemma.

If the limit is nonzero, then $\frac13\le\kai<\infty$. 
To show $\kai\ge1$, suppose otherwise. Then there is some $u_0>0$ for which 
$\kai Q(u)-1<0$ when $0<u\le u_0$. For sufficiently large $\tau$ this means
that $\D_\tau\chu<0$ if $\chu\le u_0$; thus there is a characteristic
satisfying $0<\chu(\tau,u_1)<u_0$ for $\tau$ sufficiently large.
Then \eqref{IIevol1} and the fact that $\psi(\tau,\cdot)$ is increasing
yield that $\psi(\tau,u)\ge e^\tau\psi_0(u_1)$ for $u_0\le u\le1$,
hence $\int_0^1\psi(\tau,u)\,du\ge e^\tau(1-u_0)\psi_0(u_1)\to\infty$
as $\tau\to\infty$. This contradicts volume conservation, so $\kai\ge1$.

Volume conservation also implies that $\psii(u_0)>0$ for some $u_0\in(0,1)$.
We aim to show that for any $u_1\in(0,1)$, $\psii(u_1)>0$ and
\be
\ln\psii(u_1)-\ln\psii(u_0) =\tau_{10}:= \int_{u_0}^{u_1}
\frac{du}{u(\kai Q(u)-1)}.
\label{IItosho}
\ee
First, note that if we define $\chu_\infty$ to be the solution of 
\[
\frac{\D\chu_\infty}{\D\tau}= (\kai Q(\chu_\infty)-1)\chu_\infty,
\quad \chu_\infty(0)=u_0,
\]
then $\chu_\infty(\tau)$ is defined for $\tau$ in a maximal interval
of the form $(-\infty,T_0)$, with $\chu_\infty(-\infty)=0$,
$\chu(T_0)=1$.  Substituting $u=\chu_\infty(\tau)$ into the integral
in \eqref{IItosho} reveals that $\chu_\infty(\tau_{10})=u_1$.

Consider the characteristic for \eqref{IIevol2} passing through
$u_0$ at a large time $\tau_0$, as described by $u=\chu_0(\tau)$ 
where
\[
\frac{\D\chu_0}{\D\tau} = (\ka(\tau_0+\tau)Q(\chu_0)-1)\chu_0,
\quad \chu_0(0)=u_0.
\]
Then since $\ka(\tau)\to\kai$ as $\tau\to\infty$ and
$Q$ is smooth on $(0,1)$, standard arguments using Gronwall's inequality
show that given any compact interval $[a,b]\subset (-\infty,T_0)$,
we have
\[
\sup_{a\le\tau\le b}|\chu_0(\tau)-\chu_\infty(\tau)|\to0
\mbox{\quad as $\tau_0\to\infty$.} 
\]
Fix such an interval that contains 0 and $\tau_{10}$ in its interior.
Note that it follows that $\D_\tau\chu_0>0$ for $\tau\in[a,b]$ if
$\tau_0$ is sufficiently large. 

From \eqref{IIevol1} it follows that for any fixed $\tau\in[a,b]$ we have
\[
\psi(\tau_0+\tau,\chu_0(\tau))=e^\tau\psi(\tau_0,u_0) \to
e^\tau \psii(u_0) \mbox{\quad as $\tau_0\to\infty$}.
\]
If $\tau<\tau_{10}$, then for sufficiently large $\tau_0$ we will have
$\chu_0(\tau)<\chu_\infty(\tau_{10})=u_1$, therefore
$\psi(\tau_0+\tau,\chu_0(\tau))\le \psi(\tau_0+\tau,u_1)$
and it follows $e^\tau\psii(u_0)\le\psii(u_1)$.
Similarly, if $\tau>\tau_{10}$ it follows 
$e^\tau\psii(u_0)\ge\psii(u_1)$. Therefore
$e^{\tau_{10}}\psii(u_0)=\psii(u_1)$, and this proves the claim
in \eqref{IItosho}. 

It remains to show that $\psii$ is continuous at $u=1$, so
that \eqref{IItosho} holds also with $u_1=1$.  Since 
$Q$ is increasing, for some $\delta>0$ we have $\D_\tau\chu_0>\delta$ 
whenever $u_0<\chu_0<1$ and $\tau_0$ is sufficiently large. 
Let $\eps>0$ be small. Then since $\psi(\tau,\cdot)$ is left continuous,
for any sufficiently large $\tau_0$ there exists $u_0$ close to 1 such that
$\psi(\tau_0,u_0)>\psii(1)-\eps$. With notation as above for this
different value of $u_0$ we have $\chu_0(-\eps)<1-\delta\eps$, 
and therefore
\[
\psi(\tau_0-\eps,1-\delta\eps)
\ge \psi(\tau_0-\eps,\chu_0(-\eps))
= e^{-\eps}\psi(\tau_0,u_0)
\ge e^{-\eps}(\psii(1)-\eps)
\]
The extreme members of these inequalities do not depend upon
the choice of $u_0$, so we can take $\tau_0\to\infty$ and deduce
that $\psii(1)\le e^\eps\psii(1-\delta\eps)+\eps$.
Since $\eps$ is arbitrary,
we infer that $\psii$ is continuous at $1$, as required.

From \eqref{IItosho} and the normalizations in \eqref{IIvol} and \eqref{SSvol}
it now follows that $\psii(u)=\SFP(u)$
where $p=1/(\kai-1)$, as claimed.
Since $\SFP$ is continuous and $\psi(\tau,\cdot)$ is
increasing, a simple argument using pointwise convergence on a 
dense set in $[0,1]$ shows that $\psi(\tau,u)\to\psii(u)$ uniformly in $u$.
\eproof
\end{proof}

As a consequence of Proposition \ref{P.comp} and Lemma \ref{L.lim}, we
find the following result, which implies that $\psi(\tau,u)$
fails to converge for a large set of initial data.

\begin{corollary} \label{P.nec1}
Assume that $\lim_{\tau\to\infty}\psi(\tau,u)=\psii(u)$ exists
for each $u\in[0,1]$. 
Then there exists $p\in[0,\infty]$ such that 
$\psii(u)=\SFP(u)$ and 
\[
\inf_{u>0}\psi_0(u)/u^q=0 \mbox{\quad for all $q<p$, }
\]
and
\[
\sup_{u>0}\psi_0(u)/u^q=\infty \mbox{\quad for all $q>p$.}
\]
\end{corollary}

These results imply the instability of the stationary solutions $\SFP$
with respect to the sup norm topology, say, or any topology for which
\begin{itemize}
\item[(a)] the map from $\psi(\tau,\cdot)$ to $\ka(\tau)$
is continuous for fixed $\tau$, and 
\item[(b)] arbitrarily small perturbations of $\SFP$ can yield 
initial data for which 
\be
\inf_{u>0}\frac{\psi_0(u)}{u^q}>0 \mbox{\quad for some $q<p$.}
\label{IIqp}
\ee
\end{itemize}

\begin{corollary}
Given any $p\in(0,\infty]$, the stationary solution $\SFP$ 
is unstable to perturbations that yield initial data that satisfy
\eqref{IIqp}.
In particular, the LSW solution (with $p=\infty$) is unstable
to all perturbations for which
$\inf_{u>0}\psi_0(u)/u^q>0$ for some $q<\infty$.
\end{corollary}

See section 6 for examples and further discussion.

\subsubsection{Regularly varying functions}
Corollary \ref{P.nec1}
leads one to expect that if $\psi(\tau,u)\to \SFP(u)$
for $0\le u\le1$ and if $p=1/(\ks-1)<\infty$,
then $\psi_0(u)\sim cu^p$ as $u\to0$. 
This is almost but not quite exactly correct. 
In Theorem \ref{T.nec} below, we give a
necessary criterion for convergence in this case that we conjecture is sharp. 
This criterion is related to the concept of regularly varying function 
as is treated in the books of Seneta\cite{Seneta} 
and Bingham et al.,\cite{Bing} for example.

\begin{definition} \label{D.regvar}
A positive, measurable function $g$, defined on some interval of the
form $(0,a]$, is called {\em regularly varying at $0$}
(with exponent $p\in\R$) if 
\be
\lim_{x \to 0^+} \frac{ g(\lambda x)}{\lambda^{p} g(x)} =1
\quad\mbox{ for all $\lambda>0$.}
\label{IIreg}
\ee
If $p =0$, we say $g$ is {\em slowly varying} at $0$.
\end{definition}

\begin{definition} \label{D.loclin}
{\rm
We say that a real-valued, measurable function $h$, defined on some
interval $[A,\infty)$, is {\em locally linear at $\infty$} 
(with {\it slant} $p\in\R$) if 
\be
\lim_{y \to \infty} h(y+L) -h(y) = p L \label{IIll}
\quad\mbox{ for all $L\in\R$.}
\ee
If $p=0$ we say $h$ is {\em locally flat} at $\infty$.
}
\end{definition}

It is a classic result (cf. Lemma 1.1 of ref.~\citen{Seneta} 
or Theorem 1.2.1 of ref.~\citen{Bing}) that if
$g$ is regularly varying at $0$, then the limit in \eqref{IIreg}
is attained {\it uniformly} for all values of $\lambda$ in any 
compact interval $[a,b]\subset(0,\infty)$.
Similarly, if
$h$ is locally linear at $\infty$, then the limit in  \eqref{IIll}
is achieved {uniformly} for all values of $L$ in any fixed finite interval.
This fact implies we have the following characterization of the 
two definitions just given. We define the {\it oscillation} of a function $h$
on an interval $[a,b]$ to be
\[
\osc_{z\in[a,b]} h(z) = \sup_{z\in[a,b]} h(z) - \inf_{z\in[a,b]} h(z)
= \sup_{z_1,z_2\in[a,b]} h(z_1)-h(z_2).
\]

\begin{lemma} \label{L.reg1}
Suppose $h(-\ln x)=-\ln g(x)$. Then the following are equivalent.
\begin{itemize}
\item[(i)] $g$ is regularly varying at 0 with exponent $p$.
\item[(ii)] $h$ is locally linear at $\infty$ with slant $p$.
\item[(iii)] For all $L>0$, 
\[
\osc_{z\in[y,y+L]} (h(z)-pz) \to 0
\mbox{\quad as $y\to\infty$.} 
\]
\item[(iv)] For all $\lambda>1$,
\[
\frac{ \sup_{z\in[x,\lambda x]} g(z)/z^p}
{ \inf_{z\in[x,\lambda x]} g(z)/z^p}
\to 1 \mbox{\quad as $x\to0$}.
\]
\end{itemize}
\end{lemma}

The property of being regularly varying is invariant under
a ``nice'' change of variables. 

\begin{lemma} \label{L.reg2}
Suppose $x\mapsto X(x)$ is $C^1$ near $x=0$ with
$X(0)=0$, $X'>0$. Then, $g$ is regularly varying at 0 with 
exponent $p$ if and only if $g\circ X$ has the same property.
\end{lemma}

\begin{proof}
It suffices to prove the ``only if'' part, since we can write
$g=(g\circ X)\circ X^{-1}$ and $X^{-1}$ has all the properties
listed above for $X$.  Assume that $g$ is regularly varying at 0
with exponent $p$, and let $a<b$ with $[a,b]\subset(0,\infty)$.
If $\lambda\in(a,b)$ is fixed
we have 
\[
\tilde\lambda(x):=\lambda\frac{ X(\lambda x)}{\lambda x}
\frac{x}{X(x)}\to\lambda
\]
as $x\to0$, hence $\tilde\lambda(x)\in(a,b)$ for sufficiently
small $x$. Then, from the fact that \eqref{IIreg} holds
uniformly for $\lambda\in[a,b]$ we infer that
\[
\frac{g(X(\lambda x))}{\lambda^p g(X(x))}
= \frac{ g(\tilde\lambda(x)X(x))}{\tilde\lambda(x)^pg(X(x))}
\left(\frac{\tilde\lambda(x)}{\lambda}\right)^p \to 1
\mbox{\quad as $x\to0^+$.}
\]
This proves that $g\circ X$ is regularly varying at $0$ with
exponent $p$.
\eproof
\end{proof}

\subsubsection{Convergence requires regularly varying data}
Recall from \eqref{IIevol1} that 
\[
\chu(\tau,u)=1-\frac{\chv(t,e^\tau\bar v(0)(1-u))}{e^{\tau}\bar v(0)}
\]
where $\tau=\ln(\bar v(t)/\bar v(0))$, and that $\chv(t,\cdot)$ is analytic
in a neighborhood of $\bar v(t)$ with positive derivative.
From Lemma \ref{L.reg2} and formula \eqref{IIevol1}, we find the following.

\begin{lemma} \label{L.reg3}
For any $\tau>0$ and $p\in[0,\infty)$, 
$\psi(\tau,\cdot)$ is regularly varying at 0 with exponent $p$
if and only if $\psi_0$ has the same property.
\end{lemma}

Here is our main result that gives a necessary condition for 
convergence in the rescaled variables to any one of the stationary
solutions $\SFP$ with $0\le p<\infty$.
(This result does not address the case in which the limit is
the LSW-solution, the case $p=\infty$ in Lemma \ref{L.sim}.) 

\begin{theorem} \label{T.nec}
Assume that for some $p\in[0,\infty)$ we have
\[
\lim_{\tau\to\infty}\psi(\tau,u)=\SFP(u)
\quad\mbox{ for all $u\in[0,1]$.}
\]
Then $\psi_0$ is regularly varying at $0$ with exponent $p$.
\end{theorem}

\begin{proof}
By Lemma \ref{L.reg3}, to prove the theorem 
it suffices to show that for some $\tau_0>0$,
$\psi(\tau_0,\cdot)$ is regularly varying at $0$ 
with exponent $p$. We consider first the case $p>0$.
Since $\ka(\tau)\to\ks$ as $\tau\to\infty$, we may fix
$\tau_0>0$ so that $|\ka(\tau)-\ks|\le 1/2p$ for all $\tau\ge\tau_0$.

In the analysis to come we shall often make use of the variables
\be
y=-\ln u, \quad \lnp= -\ln\psi.
\label{IIvar2}
\ee
With $\lnp_0(y)=-\ln\psi_0(u)$, we have
\be
\lnp(\tau,\chy(\tau,y))= \lnp_0(y)-\tau
\label{IIevol3}
\ee
whenever $\chy(\tau,y)>0$, where $\chy=-\ln\chu$ satisfies
\be
\frac{\D\chy}{\D\tau}= -\ka(\tau)Q(e^{-\chy})+1 ,
\quad \chy(0,y)=y.
\ee
For differentiable initial data,
\be
\D_\tau\lnp -(\ka(\tau)Q(e^{-y})-1)\D_y\rho = -1.
\label{IIevol4}
\ee

In these variables stationary solutions are described 
via $\rs(y)=-\ln\SF(u)$, and we have
\be
(\ks Q(e^{-y})-1)\rs'(y)=1
\label{IIrseq}
\ee
for all $y>0$. For $0<p=1/(\ks-1)<\infty$ it follows
\be
\rs(y) = p y - A_*(y) 
\label{IIAs}
\ee
where $\lim_{y\to\infty}A_*(y)$ exists (and therefore
$A_*$ is locally flat at $\infty$).

Define
\be
h(\tau,y)= \rho(\tau,y)-\rs(y).
\label{IIhdef}
\ee
The uniform convergence established in Lemma \ref{L.lim}
implies that for any compact interval $[a,b]\subset[0,\infty)$,
$\lim_{\tau\to\infty} h(\tau,y)=0$ uniformly for $y\in[a,b]$. 

To prove the theorem, it is enough to show that $h(\tau_0,\cdot)$ is
locally flat at $\infty$. This will be done by following
pairs of characteristics backward to time $\tau_0$ from large times
$\ttau$. The idea in the estimates that follow is that 
the total change in distance between such pairs can be bounded
independently of $\ttau$.

Let $\tilde\chy(\tau,\ty)$ denote the characteristic satisfying 
\be\label{IIYdef}
\frac{\D\tilde\chy}{\D\tau} = -(\ka(\tau)Q(e^{-\tilde\chy})-1),
\quad\tilde\chy(\ttau,\ty)=\ty.
\ee
For $\ttau\ge\tau\ge\tau_0$ we have
\be\label{IIdyest}
- \frac{\D\tilde\chy}{\D\tau} \ge \ka(\tau)-1\ge \frac1{2p}
\ee
hence
\be
\tilde\chy(\tau,\ty) \ge \ty+ (\ttau-\tau)/2p.
\label{IIyest}
\ee
We also note that $\ka(\tau)\le K_0:=\ks+1/2p$ for $\tau\ge\tau_0$,
so $\tilde\chy(\tau,\ty)\le \ty+3K_0(\ttau-\tau)$.

Since $Q'(0)=1/3$, there is some $\delta_0>0$ such that 
$Q'(e^{-y})<1$ whenever $y\ge\delta_0$. 
Suppose that $\delta_0\le\ty\le\min(\ty_1,\ty_2)$. Let
\be \begin{array}{ll}
\tchy_1(\tau)=\tchy(\tau,\ty_1), & y_1=\tchy_1(\tau_0),\\
\tchy_2(\tau)=\tchy(\tau,\ty_2), & y_2=\tchy_2(\tau_0),
\end{array}
\ee
We compute that
\be\label{IIEdef}
E(\ttau,\ty_1,\ty_2):=
(y_1-y_2) - (\ty_1-\ty_2) = 
\int_{\tau_0}^{\ttau} \ka(\tau)(
Q(e^{-\tchy_1(\tau)})-
Q(e^{-\tchy_2(\tau)}) )\,d\tau.
\ee
Using \eqref{IIyest} and the estimate
$| Q(e^{-\tchy_1})- Q(e^{-\tchy_2})|\le e^{-\tchy}$, we find that
\be
|E(\ttau,\ty_1,\ty_2)|\le K_0 e^{-\ty}
\int_{\tau_0}^{\ttau} e^{-(\ttau-\tau)/2p}d\tau
\le 2pK_0 e^{-\ty}.
\label{IIEest}
\ee
From the relations \eqref{IIhdef}, \eqref{IIevol3} and \eqref{IIrseq} we find that
\begin{eqnarray}
H(\ttau,\ty_1,\ty_2)&:=&
(h(\tau_0,y_1)-h(\tau_0,y_2)) - 
(h(\ttau,\ty_1)-h(\ttau,\ty_2))  \nonumber
\\&=& \int_{\tau_0}^{\ttau} (\ks-\ka(\tau)) \frac
{ Q(e^{-\tchy_2(\tau)})- Q(e^{-\tchy_1(\tau)}) }
{ (\ks Q(e^{-\tchy_1(\tau)})-1) (\ks Q(e^{-\tchy_2(\tau)})-1)}
\,d\tau.
\label{IIhdif}
\end{eqnarray}
As above we have the estimate
\be
|H|\le M_0(\ttau):= p^2 e^{-\ty} \int_{\tau_0}^{\ttau}
|\ks-\ka(\tau)| e^{-(\ttau-\tau)/2p}\,d\tau.
\label{IIHest}
\ee
Since $\ka(\tau)-\ks\to0$, it follows $M_0(\ttau)\to0$ 
as $\ttau\to\infty$.

Let $L>0$, and fix an arbitrary $\ty>\delta_0$. 
Put $\ty_2=\ty$ and choose
$\ty_1>\ty_2$ (depending on $\ttau$) so that with 
$y=y(\ttau)=y_2$ we have $y+L=y_1$. 
By \eqref{IIEest} we have the bound $\ty_1-\ty_2\le L+2pK_0$,
so we may deduce that $h(\ttau,\ty_1)\to0$ as $\ttau\to\infty$.

Taking $\ttau\to\infty$ in \eqref{IIhdif}, we find that
\be
\lim_{y\to\infty} h(\tau_0,y+L)-h(\tau_0,y) = 0.
\label{IIoest}
\ee
Since $L>0$ is arbitrary, 
it follows that $h(\tau_0,\cdot)$ is locally flat at $\infty$. 
This finishes the proof of the theorem in the case $p>0$.

It remains to consider the case $p=0$, when $\SFP(u)=1$ for $u\in(0,1]$.
In this case $\rs=0$ and in \eqref{IIhdif} we have $H(\ttau,\ty_1,\ty_2) = 0$.
Moreover, $\lim_{\tau\to\infty}\ka(\tau)=\infty$. We fix $\tau_0$ sufficiently
large so that $\ka(\tau)>2$ for $\tau\ge\tau_0$.

The strategy of the proof is the same as above. The key is to show that
$E(\ttau,\ty_1,\ty_2)$ is bounded independent of $\ttau$ for $\ttau>\tau_0$.
Starting from \eqref{IIYdef},
the first inequality in \eqref{IIdyest} remains valid, and moreover yields
$\ka(\tau)\le -2{\D\tilde\chy}/{\D\tau}$ for $\tau>\tau_0$. From \eqref{IIEdef}
we get the bound
\[
|E(\ttau,\ty_1,\ty_2)| \le 
\int_{\tau_0}^\ttau \ka(\tau) e^{-\tchy(\tau)}d\tau \le 
2\int_{\tau_0}^\ttau e^{-\tchy(\tau)}(-{\D\tilde\chy}/{\D\tau})d\tau \le 2.
\]
Arguing as before, we obtain \eqref{IIoest} and the claimed result follows.
\eproof
\end{proof}

\subsection{A sufficient condition for sufficiency}

The estimates in the preceding proof can be used to establish some
other interesting results. We conjecture that the condition that
$\psi_0$ is regularly varying at 0 with exponent $p\ge0$ is actually
sufficient as well as necessary for convergence to 
$\SFP$ as $\tau\to\infty$. A general proof is lacking, 
but in the subsection to follow, sufficiency is proved when 
$p>0$ is sufficiently small and the initial data
is close to $\SFP$ in a sense related to the notion of regularly varying
functions.
The result given here reduces the general problem of sufficiency to proving 
the convergence of $\ka(\tau)$.

\begin{theorem} \label{T.cconv}
Assume $\lim_{\tau\to\infty}\ka(\tau)=\ks\in(1,\infty]$. 
Then, $\lim_{\tau\to\infty}\psi(\tau,u)$ exists for all $u\in[0,1]$
if and only if $\psi_0$ is regularly varying with exponent $p=1/(\ks-1)$.
\end{theorem}

\begin{proof}
The ``only if'' part has already been proved. Suppose $\psi_0$ is
regularly varying with exponent $p$. We first consider the case $p>0$.
Then for $q<p$ we have 
$\psi_0(u)/u^q\to 0$ as $u\to0$ and for $q>p$ we have
$\psi_0(u)/u^q\to \infty$ as $u\to0$. 
Since we assume $\ka(\tau)\to\ks$ as $\tau\to\infty$,
Proposition \ref{P.comp} implies that $p=1/(\ks-1)$.
For sufficiently large $\tau_0$ we have $\ka(\tau)-1\ge 1/2p$ 
whenever $\tau\ge\tau_0$,
and we can use the estimates from the proof of Theorem \ref{T.nec}.

Our first step is to show that for any $L>0$,
\be
\sup_{z\in[\delta_0,\delta_0+L]} |h(\ttau,z)-b(\ttau)|\to0
\quad\mbox{ as $\ttau\to\infty$,}
\label{CCclaim1}
\ee
where we define $b(\ttau)=h(\ttau,\delta_0)$.
To prove this we use \eqref{IIhdif}, taking $\ty_2=\delta_0$ and
$\ty_1\in[\delta_0,\delta_0+L]$. Then $\tchy_1\ge\tchy_2$ and it follows
that $E(\ttau,\ty_1,\ty_2)\le0$ (the characteristics are diverging),
so we have
\[
\delta_0+(\ttau-\tau_0)/2p\le y_2\le y_1\le y_2+L.
\]
Now \eqref{IIhdif} yields that, 
\[
|h(\ttau,\ty_1)-b(\ttau)|\le \osc_{z\in[y_2,y_2+L]}h(\tau_0,z)+M_0(\ttau)
\to0
\]
as $\ttau\to\infty$ uniformly for $\ty_1\in [\delta_0,\delta_0+L]$.
This proves \eqref{CCclaim1}.

Next, we show that for any $L>\delta_0$, 
\be
\sup_{z\in[0,L]} |h(\ttau,z)-b(\ttau)| \to 0
\quad\mbox{ as $\ttau\to\infty$.}
\label{CCclaim2}
\ee
To prove this, we use \eqref{IIyest} taking $\ty=0$, and use
the identity \eqref{IIhdif} replacing $\tau_0$ by $\tau_1=\ttau-2p\delta_0$,
which guarantees that $\tchy(\tau_1,\ty)\ge\delta_0$. 
Fix $\ty_2=\delta_0$. For any $\ty_1\in[0,L]$ we have 
\[
\delta_0\le y_1\le \tilde L:=L+3K_0(\ttau-\tau_1) = L+6p\delta_0K_0.
\]
Then since $1\le Q\le3$ it follows from \eqref{IIhdif} that
\[
|h(\ttau,\ty_1)-b(\ttau)| \le \osc_{z\in[\delta_0,\tilde L]}h(\tau_1,z)
+3p^2 \int_{\tau_1}^\ttau |\ks-\ka(\tau)|\,d\tau 
\]
and this tends to zero as $\ttau\to\infty$ uniformly for $\ty_1\in[0,L]$,
proving \eqref{CCclaim2}.

Now \eqref{CCclaim2} implies that for any $L>0$, $\rho(\tau,y)-b(\tau)\to\rs(y)$
as $\tau\to\infty$ uniformly for $y\in[0,L]$. Since 
$\rho(\tau,y)-b(\tau)$ increases to $\infty$ as $y\to\infty$ for each $\tau$,
we infer from the volume conservation identity \eqref{IIvol} that
as $\tau\to\infty$,
\[
e^{b(\tau)}= \int_0^\infty 
e^{-\rho(\tau,y)+b(\tau)} e^{-y}\,dy \to
\int_0^\infty 
e^{-\rs(y)} e^{-y}\,dy =1.
\]
Hence $\lim_{\tau\to\infty}b(\tau)=0.$
Then it quickly follows that $\lim_{\tau\to\infty}\psi(\tau,u)$
exists for each $u\in[0,1]$. This finishes the proof when $p>0$.

In the remaining case we assume $\lim_{\tau\to\infty}\ka(\tau)=\infty$
and $\psi_0$ is regularly varying with exponent $p=0$. 
Then $h(\tau_0,\cdot)=\rho(\tau_0,\cdot)$ is locally flat at $\infty$ for any $\tau_0\ge0$. 
We take
$\tau_0$ sufficiently large so that $\ka(\tau)>2$ for $\tau>\tau_0$,
and take $\ty_2=0$ and $\ty_1\in[0,L]$. Since 
$\tchy_1(\tau)\ge\tchy_2(\tau)\ge \ttau-\tau$ 
for $\tau_0\le\tau\le\ttau$, it follows $E(\ttau,\ty_1,\ty_2)\le0$, hence
$\ttau-\tau_0\le y_2\le y_1\le y_2+ L$. 
Since $H=0$ in \eqref{IIhdif}, we infer directly that
\[
\sup_{z\in[0,L]} |h(\ttau,z)-h(\ttau,0)| \le \osc_{z\in[y_2,y_2+L]}h(\tau_0,z) \to0
\]
as $\ttau\to\infty$. With $b(\tau)=h(\tau,0)$, this means that for all $L>0$,
$\rho(\tau,y)-b(\tau)\to0$ as $\tau\to\infty$ uniformly for $y\in[0,L]$.
As above, from volume conservation we deduce that $b(\tau)\to0$ and
hence $\lim_{\tau\to\infty}\psi(\tau,u)=1$ for all $u\in(0,1]$. 
\eproof
\end{proof}

\subsection{Stability and convergence for small $p$}

In trying to establish convergence of $\psi(\tau,u)$ as $\tau\to\infty$
for regularly varying initial data,
the main obstacle is achieving a priori control over $\ka(\tau)-\ks$.
We can get such control when $p$ is small, and obtain a stability
result for the corresponding stationary solutions $\SFP$, with
respect to perturbations that are small in a sense related to 
the notion of regularly varying functions. 

For perturbations that are small in the appropriate sense, 
the condition that the initial data is
regularly varying is necessary and sufficient for 
convergence as $\tau\to\infty$. In particular, the 
result below implies convergence to  $\SFP$ (for $p>0$ small)
for some initial data for which $\lim_{u\to0}\psi_0(u)/u^p$ does not exist.
If the initial data are not regularly varying, convergence 
cannot occur.

Recall that 
\[
h(\tau,y)=\rho(\tau,y)-\rs(y) = -\ln(\psi(\tau,e^{-y})/\SFP(e^{-y})).
\]
We define the {\em flatness modulus} of $h$ to be
\be
\mf(\tau,y) = \sup_{\ty\ge y} \osc_{z\in[\ty,\ty+1]} h(\tau,z).
\label{CCmf}
\ee
Note that whenever $0\le a<b<c$, 
\[
\osc_{z\in[a,c]}h(\tau,z) \le
\osc_{z\in[a,b]}h(\tau,z) +
\osc_{z\in[b,c]}h(\tau,z),
\]
so for any positive integer $n$, 
\be
\osc_{z\in[y,y+n]} h(\tau,z)\le n\mf(\tau,y).
\label{CCosc}
\ee
Evidently, $\mf(\tau,y)$ is decreasing in $y$,
and we have $\mf(\tau,y)\to0$ as $y\to\infty$
if and only if $h(\tau,\cdot)$ is locally flat at $\infty$.

\begin{theorem} \label{T.stable}
For sufficiently small $p>0$, there exist positive constants
$\delta_*$ and $K_0$ with the following property. 
If the flatness modulus of $h$ satisfies $\mf(0,0)\le\delta_*$,
then 
\be
\mf(\tau,0)\le K_0\mf(0,0)
\label{mfest}
\ee
and
\be
\sup_{0<u\le1}|\psi(\tau,u)-\SFP(u)|\le K_0\mf(\tau,0)
\label{Stest}
\ee
for all $\tau\ge0$. 

If in addition 
$\psi_0$ is regularly varying at 0 with exponent $p$
(equivalently, if $h(0,\cdot)$ is locally flat at $\infty$),
then $\mf(\tau,0)\to0$ as $\tau\to\infty$, and 
$\lim_{\tau\to\infty}\psi(\tau,u)=\SFP(u)$ for all
$u\in[0,1]$.
\end{theorem}

\begin{proof} We start by getting a bound on $|\ka(\tau)-\ks|$
where $\ks=1+1/p$.
From \eqref{IIka}, using $\SF=\SFP$ we have the representation
\begin{eqnarray}
\frac{1}{\ka(\tau)}- \frac{1}{\ks} &=& 
\int_0^1 \left(
\frac{\SF(u)}{\SF(1)}-\frac{\psi(\tau,u)}{\psi(\tau,1)}
\right)(1-u)^{-2/3}\,du \nonumber
\\&=& \int_0^\infty \left(
1-e^{h(\tau,0)-h(\tau,y)}\right) e^{\rs(0)-\rs(y)}(1-e^{-y})^{-2/3}e^{-y}\,dy.
\label{CCkdif}
\end{eqnarray}
We know that $\rs(y)-\rs(0)\ge0$, and \eqref{CCosc} implies
\be\label{hest}
|h(\tau,0)-h(\tau,y)|\le (y+1)\mf(\tau,0).
\ee
So, since $|1-e^x|\le|x|e^{|x|}$ for all $x$, 
the integrand in \eqref{CCkdif} is bounded by 
\[
\mf(\tau,0)(y+1)e^{\mf(\tau,0)(y+1)}(1-e^{-y})^{-2/3}e^{-y}.
\]
Thus there is a constant $\KK$ independent of $h$ and $p$
such that if $\mf(\tau,0)\le\frac12$ then
\be
\left|\frac{1}{\ka(\tau)}- \frac{1}{\ks}\right| \le
\KK \mf(\tau,0).
\label{CCkest}
\ee
It follows that if $\KK\mf(\tau,0)\le 1/4p\ks^2=p/4(p+1)^2$, then 
$\ka(\tau)\le2\ks$ and 
\be
|\ks-\ka(\tau)|\le 2\ks^2 \KK\mf(\tau,0)\le \frac1{2p}.
\label{CCkest2}
\ee

Next, we develop some a priori estimates for the flatness modulus, 
assuming it is small on some time interval.

\begin{lemma} \label{L.ker}
For $p>0$ there is a positive decreasing
function $\GP\colon[0,\infty)\to\R$ such that
\[
\int_0^\infty \GP(\tau)\,d\tau\le \CB p(p+1)^2
\]
for some constant $\CB>0$ independent of $p$, and the following holds.
Assume that for some $\tau_*>0$ we have $\KK\mf(\ttau,0)\le p/4(p+1)^2$
for $0\le\ttau\le\tau_*$. Then 
\[
\mf(\ttau,0)\le \mf(0,\ttau/2p)+ \int_0^\ttau 
\GP(\ttau-\tau)\mf(\tau,0)\,d\tau
\]
for $0\le\ttau\le\tau_*$. 
\end{lemma}

\begin{proof}
We adopt the notation of the proof of Theorem \ref{T.nec}.
First, let $\ty\ge0$ be arbitrary and suppose $\ty_1$, 
$\ty_2\in[\ty,\ty+1]$. We use the identity \eqref{IIhdif}
and the fact that $1\le Q\le3$. With any values of $\tau_0\le\ttau$
between 0 and $\tau_*$ we have 
\[
|h(\ttau,\ty_1)-h(\ttau,\ty_2)| \le
|h(\tau_0,y_1)-h(\tau_0,y_2)|+ 3p^2\int_{\tau_0}^\ttau
|\ks-\ka(\tau)|\,d\tau.
\]
Since the characteristics are diverging, $y_1$ and $y_2$ lie in
$[\hat y,\hat y+1]$ where 
$\hat y=\tchy(\tau_0,\ty)\ge\ty+(\ttau-\tau_0)/2p$, and so
since $p^2\ks^2=(p+1)^2$, using \eqref{CCkest2} we find
\[
\osc_{z\in[\ty,\ty+1]} h(\ttau,z) \le
\osc_{z\in[\hat y,\hat y+1]} h(\tau_0,z)+ 
6(p+1)^2 \KK \int_{\tau_0}^\ttau \mf(\tau,0)\,d\tau.
\]
Taking the sup over $\ty\ge y$ it follows that 
\be
\mf(\ttau,y)\le \mf(\tau_0,y+(\ttau-\tau_0)/2p)
+ 6(p+1)^2 \KK \int_{\tau_0}^\ttau \mf(\tau,0)\,d\tau.
\label{CCmf1}
\ee

Next we use the same identity \eqref{IIhdif} 
but assume that $\ty\ge\delta_0$ 
as chosen in the proof of Theorem \ref{T.nec} (so $Q'(e^{-y})<1$ for
$y>\delta_0$). Taking $\tau_0=0$, we obtain the estimate
\[
|h(\ttau,\ty_1)-h(\ttau,\ty_2)| \le
|h(0,y_1)-h(0,y_2)|+ p^2e^{-\ty} \int_0^\ttau
|\ks-\ka(\tau)|e^{-(\ttau-\tau)/2p}\,d\tau.
\]
From this we deduce that for $y\ge\delta_0$,
\be
\mf(\ttau,y)\le \mf(0,y+\ttau/2p)+ 2(p+1)^2e^{-\delta_0} \KK
\int_0^\ttau \mf(\tau,0) e^{-(\ttau-\tau)/2p}\,d\tau.
\label{CCmf2}
\ee

Now, we invoke \eqref{CCmf1} with $y=0$ and 
$\tau_0=\max(0,\ttau-2p\delta_0)$, and replace $\ttau$ in \eqref{CCmf2}
by this $\tau_0$ and use $y=\delta_0$.  Then,  with
\[
\GP(\tau)=\cases{
6(p+1)^2 \KK, & $0\le\tau\le 2p\delta_0$,\cr
2(p+1)^2 \KK e^{-\tau/2p}, & $\tau\ge 2p\delta_0$,
}
\]
we find that the 
inequalities asserted in the Lemma are valid.
\eproof
\end{proof}

Now we proceed with the proof of Theorem \ref{T.stable}.
Assume $p$ is sufficiently small so that $\CB p(p+1)^2\le\frac12$,
and assume $\KK\mf(0,0)< p/8(p+1)^2$. Let 
$M(\ttau)=\sup_{0\le\tau\le\ttau}\mf(\tau,0)$, then using Lemma
\ref{L.ker} and a simple continuation argument, we find that
$M(\ttau)\le 2\mf(0,0)$ for all $\ttau\ge0$. Thus we may take
$\delta_*= p/8\KK(p+1)^2$ and $K_0=2$ to prove the first estimate
\eqref{mfest} asserted in the Theorem.

The last part of the Theorem follows from Lemma \ref{L.ker} 
and the lemma below, together with the fact that if
$h(0,\cdot)$ is locally flat at $\infty$, 
then $\mf(0,y)\to0$ as $y\to\infty$. 
One deduces that $\mf(\tau,0)\to0$ as $\tau\to\infty$. 
This implies $\ka(\tau)\to\ks$, and the convergence of $\psi(\tau,u)$ 
follows from Theorem \ref{T.cconv}.

\begin{lemma} 
Suppose $f$, $a$, and $g$ are positive bounded measurable functions 
on $(0,\infty)$ and $\int_0^\infty g(s)\,ds=\gamma<1$. Assume
\[
f(t)\le a(t)+\int_0^t g(t-s)f(s)\,ds
\]
for all $t\ge0$, and $a(t)\to0$ as $t\to\infty$. Then $f(t)\to0$
as $t\to\infty$.
\end{lemma}

\begin{proof}
Let $F(t)=\sup_{0\le s\le t}f(s)$, $A=\sup_{t\ge0}a(t)$.
Then $F(t)\le A+\gamma F(t)$ so $F(t)\le A/(1-\gamma)$ for all $t\ge0$.
Let $\Fi=\limsup_{t\to\infty}f(t)$, and let $\eps>0$. Choose $t_0$ so
$f(s)\le\Fi+\eps$ whenever $s\ge t_0$. Then for $t\ge t_0$,
\[
f(t)\le a(t)+ \frac{A}{1-\gamma}\int_0^{t_0}g(t-s)\,ds+
\gamma(\Fi+\eps).
\]
Taking the lim sup as $t\to\infty$, we find $\Fi\le \gamma(\Fi+\eps)$.
Since $\eps$ is arbitrary, it follows $\Fi=0$.
\eproof
\end{proof}

It remains to prove the stability estimate \eqref{Stest}. First, we get a bound
on $|h(\tau,0)|$. Since $\int_0^1\psi(\tau,u)\,du=\int_0^1\SFP(u)\,du=1$ we find
\[
e^{h(\tau,0)}-1 = \int_0^\infty \left(
e^{h(\tau,0)-h(\tau,y)}-1\right) \SFP(e^{-y}) e^{-y}\,dy.
\]
Since $\SFP(u)\le \alpha_p(0)u^p$ and $|e^x-1|\le|x|e^{|x|}$ it follows from
\eqref{hest} that provided $\mf(\tau,0)\le\frac12$ we have
\[
\left|e^{h(\tau,0)}-1\right| \le \alpha_p(0) \mf(\tau,0) \int_0^\infty
(y+1)e^{(y+1)\mf(\tau,0)} e^{-(p+1)y}\,dy
\le K_1 \mf(\tau,0)
\]
where $K_1$ 
is a constant. Provided we can guarantee $\mf(\tau,0)\le 1/2K_1$, we can
ensure that $e^{h(\tau,0)}\ge \frac12$ and then
\be\label{h0est}
|h(\tau,0)| = |\ln e^{h(\tau,0)} - \ln e^0| \le 2K_1\mf(\tau,0).
\ee

From \eqref{hest} we then infer $|h(\tau,y)|\le (y+K_2)\mf(\tau,0)$ where
$K_2=2K_1+1$. Then it follows that for $u=e^{-y}\in(0,1]$, 
\be\label{psest}
|\psi(\tau,u)-\SFP(u)| \le \left|e^{-h(\tau,y)}-1\right| \SFP(e^{-y}) \le
\alpha_p(0) |h(\tau,y)|e^{|h(\tau,y)| -py}
\le K_3 \mf(\tau,0)
\ee
where $K_3$ is a constant, provided we can guarantee that $\mf(\tau,0)\le p/2$.
By \eqref{mfest}, the bounds on $\mf(\tau,0)$ that are needed to obtain \eqref{psest} 
will hold if $\mf(0,0)\le\delta_*$ for a sufficiently small value of $\delta_*$.
This finishes the proof of the Theorem.
\eproof
\end{proof}

\section{Discussion}

To recap and explain the main results of the last section, recall that 
$u=1-v/\bar v(t)$ and $\psi=(\bar v(t)/V)\F$ 
where $\F(t,x)=\int_x^\infty f(t,v)\,dv$, and suppose the initial data for $\psi$ 
is written in the form
\be \label{psi0}
\psi_0(u) = \left(\frac{\bar v_0}{V}\right) \F_0(\bar v_0(1-u)) 
= \SFP(u)e^{-h_0(\ln(1/u))}.
\ee
Here $\psi_0$ must be increasing and is normalized so $\int_0^1\psi_0(u)\,du=1$.
The results of section 5 show that:
\smallskip

\begin{itemize}
\item[(i)] If the scaled distribution function $\psi$ converges pointwise
as $\tau\to\infty$, then the limit is $\SFP$ for some $p\in[0,\infty]$. 
\item[(ii)] If $\psi_0(u)\ge au^q>0$ for some $q<\infty$ and $a>0$, 
then $\psi$ cannot converge to $\SFI$, the solution favored by the LSW theory.
\item[(iii)] If $\psi$ converges to $\SFP$ for some $p\in[0,\infty)$, 
then $\psi_0$ is regularly varying at $0$ with exponent $p$.
\item[(iv)] If $p$ is sufficiently small and positive, and if the flatness modulus
of $h_0$ is sufficiently small, then $\psi$ converges to $\SFP$ 
{\it if and only if} $\psi_0$ is regularly varying at $0$ with exponent $p$.
\end{itemize}

Recall that $\psi_0$ is regularly varying at $0$ with exponent $p$ if and only if 
the function $h_0$ in \eqref{psi0} is locally flat at $\infty$. 
($\SFP$ can be replaced by $u^p$ in \eqref{psi0} and this statement still holds.)
Thus, if $h_0$ is bounded (implying $\psi_0(u)\ge au^p$) 
and is {\it not} locally flat at $\infty$,
then $\psi$ cannot converge pointwise to {\it any} limit as $\tau\to\infty$.
For a concrete example, any periodic or quasiperiodic function $h_0$ will suffice,
e.g., $h_0=a-b\sin y$, corresponding to $\psi_0(u)=Cu^pe^{b\sin\ln(1/u)}.$

If $h_0$ is $C^1$, the mean value theorem implies that the flatness modulus
of $h_0$ satisfies
\be
\mf_0(y):= \sup_{\ty\ge y} \osc_{z\in[\ty,\ty+1]} h_0(z) \le 
\sup_{\ty\ge y}|h_0'(\ty)|,
\label{CCmf0}
\ee
and so $h_0$ is locally flat at $\infty$ if $h_0'(y)\to0$ as $y\to\infty$. 
As an example, suppose 
$h_0(y)=-\ln(1+\eps y)+c$, so $\psi_0(u)=C\SFP(u)(1+\eps\ln(1/u))$.
Then $h_0$ is locally flat at $\infty$ and $\mf_0(0)\le\eps$.
So if $p$ is small and $\eps$ is small, the corresponding 
solution $\psi$ converges to $\SFP$ in this case,
despite the fact that $\psi_0(u)/u^p\to\infty$ as $u\to0$.

Sharp representation formulae are known for regularly varying functions. 
For example, from Lemma 1.3 of ref.~\citen{Seneta}, it follows immediately that
$h_0$ is locally flat at $\infty$ if and only if
there exists $b>0$ such that for $y>b$,
\[
h_0(y)=f_1(y)+\int_b^y f_2(t)\,dt,
\]
where $f_1$ and $f_2$ are bounded measurable functions such that
$\lim_{y\to\infty}f_1(y)$ exists and $\lim_{y\to\infty} f_2(y)=0$.

The results we have established in this paper show that the 
asymptotic behavior of solutions is
exquisitely sensitive to the behavior of the initial data near the tip of the support. 
The conditions in (ii) and (iii) 
that ensure nonconvergence in the rescaled variables 
depend only on the initial size distribution of 
an arbitrarily small fraction of the largest particles. 
That is, they depend only on the initial volume ranking $v_0(\F)$
for values of the particle rank $\F$ in an arbitrarily small interval $(0,\eps_0)$.

It is clear that one can always perturb the sizes of
an arbitrarily small fraction of the largest particles, 
by a maximum amount as small as you like,
to guarantee that $\psi_0$ is not regularly varying at 0 
and $\psi_0(u)\ge au^p$ for some $p$.
This means that, with respect to the topology given by the sup norm
distance between volume rankings (a natural topology for well-posedness
of the initial value problem), 
there is a dense set of initial data that yield nonconvergence to 
self-similar form. 

By similarly perturbing the sizes of the largest particles, one can guarantee that
$\psi_0$ is regularly varying at $0$ with any given exponent $p$.
If it is true, as we conjecture, that having such data implies that
$\psi$ converges to the self-similar profile $\SFP$, then there is a dense set
of initial data that yield convergence to $\SFP$, for any $p\in[0,\infty)$.

One can also make an arbitrarily small perturbation 
to make the new volume ranking constant on a small
interval of the form $(0,\eps)$. This means that the size distribution 
contains a Dirac delta at the tip, so by Proposition~\ref{P.atom}
the perturbed size distribution will converge to 
a stationary Dirac delta as $t\to\infty$.

We have not discussed in this paper any conditions on initial data
for convergence to the smooth self-similar profile $\SFI$ 
favored by the LSW theory.
Lemma~\ref{L.lim} implies that $\ka(\tau)\to1$ is necessary for such 
convergence.
In this case, characteristics near the tip spread more slowly and
different sorts of estimates seem to be needed to study the 
convergence question.

\bigskip\noindent{\bf Acknowledgments.}
The authors are grateful for discussions concerning this work
with S. M\"uller, F. Otto, J. Kristensen, G. Friesecke, and R. Kohn. 
Part of this work was performed at the
Max-Planck Institute for Mathematics in the Sciences in Leipzig.
This work was partly supported by the National Science Foundation
under grant DMS-9704924, and the 
Deutsche Forschungsgemeinschaft through the SFB 256.

\end{spacing}
\begin{spacing}{1}

\end{spacing}

\begin{thebibliography}{99}
\frenchspacing
\bibitem{BF} C. Berg and G. Forst,
{\it Potential Theory on Locally Compact Abelian Groups},
Springer-Verlag, New York, 1975.

\bibitem{Bing} N. H. Bingham, C. M. Goldie and J. L. Teugels,
{\it Regular Variation}, 
Encycl. Math. Appl. v. 27,
Cambridge Univ. Press, Cambridge, 1987.

\bibitem{B1} L. C. Brown,
{A new examination of classical coarsening theory,}
{\it Acta Metall. \bf37:}71--77 (1989).

\bibitem{B2} L. C. Brown,
{Reply to further comments by Hillert, Hunderi and Ryum on 
``A new examination of classical coarsening theory,"}
{\it Scripta Metall. Mater. \bf26:}1939--1942 (1992).

\bibitem{B3} L. C. Brown, 
{A new examination of volume fraction effects during particle coarsening,}
{\it Acta Metall.\ Mater. \bf40:}1293--1303 (1992).

\bibitem{CP} J. Carr and O. Penrose,
{Asymptotic behavior in a simplified Lifshitz-Slyozov equation,}
to appear in {\it Physica D}.

\bibitem{CG} J.-F. Collet and T. Goudon,
{On solutions of the Lifshitz-Slyozov model,}
preprint.

\bibitem{GMS} B. Giron, B. Meerson and P. V. Sasorov,
{Weak selection and stability of localized distributions in Ostwald
ripening,} 
{\it Phys.\ Rev.\ E \bf58:}4213--6 (1998).


\bibitem{LS} I. M. Lifshitz and V. V. Slyozov,
{The kinetics of precipitation from supersaturated solid solutions,}
{\it J.\ Phys.\ Chem.\ Solids \bf 19:}35--50 (1961).

\bibitem{MS} B. Meerson and P. V. Sasorov,
{Domain stability, competition, growth and selection in globally
constrained bistable systems,}
{\it Phys.\ Rev.\ E \bf 53:}3491--4 (1996).

\bibitem{N1} B. Niethammer,
{Derivation of the LSW theory for
Ostwald ripening by homogenization methods,}
to appear in {\it Arch.\ Rat.\ Mech.\ Anal.}

\bibitem{NP1} B. Niethammer and R. L. Pego,
On the initial value problem in the Lifshitz-Slyozov-Wagner theory
of Ostwald ripening, SFB 256 preprint, University of Bonn.

\bibitem{Pen} O. Penrose,
{The Becker-D\"oring equations at large times and their connection
with the LSW theory of coarsening,}
{\it J. Stat. Phys. \bf 89:}305--320 (1997).


\bibitem{Seneta} E. Seneta, {\it Regularly Varying Functions},
{Lec.\ Notes in Math. \bf 508}, Springer-Verlag, New York, 1976.
 
\bibitem{V} P. W. Voorhees,
{The theory of Ostwald ripening,}
{\it J. Stat.\ Phys. \bf 38:}231--252 (1985).

\bibitem{V2} P. W. Voorhees, 
{Ostwald ripening of two-phase mixtures,}
{\it Ann. Rev. Mater. Sci. \bf22:}197--215 (1992).

\bibitem{Wa} C. Wagner,
Theorie der Alterung von Niederschl\"agen durch Uml\"osen,
{\it Z. Elektrochem. \bf 65:}581--594 (1961).

\end {thebibliography}
\end{document}